\documentclass[journal]{new-aiaa}
\usepackage[utf8]{inputenc}
\usepackage{textcomp}

\usepackage{graphicx}
\usepackage{amsmath}
\usepackage{booktabs}
\usepackage{siunitx}
\usepackage{longtable,tabularx}
\usepackage{tikz}
\usepackage{algorithm}
\usepackage{algpseudocode}
\usetikzlibrary{arrows.meta,decorations.pathmorphing,calc,patterns}

\newcommand{\FluxRef}{0.76}
\newcommand{\TargetInterval}{9}
\newcommand{\DiurnalRatio}{5.7}
\newcommand{\DiurnalPeak}{7.5}
\newcommand{\DiurnalMin}{18.5}
\newcommand{\SeasonRatio}{1.7}
\newcommand{\PowerExp}{0.51}
\newcommand{\HotspotOffset}{152}

\newcommand{\BestDist}{1000}
\newcommand{\PlaceHAPS}{39}
\newcommand{\PlaceOld}{56}
\newcommand{\PlaceMid}{67}
\newcommand{\HAPSSide}{516}

\newcommand{\ThrA}{7.5}

\newcommand{\DutyA}{30}
\newcommand{\RateA}{713}
\newcommand{\WaitA}{10.9}
\newcommand{\MedA}{7.9}
\newcommand{\PninetyA}{56}
\newcommand{\MedSmallA}{3.18}
\newcommand{\ThrB}{7.4}
\newcommand{\GainB}{1.00}
\newcommand{\DutyB}{31}
\newcommand{\RateB}{824}
\newcommand{\WaitB}{24.5}
\newcommand{\MedB}{7.1}
\newcommand{\PninetyB}{111}
\newcommand{\MedSmallB}{2.74}
\newcommand{\ThrC}{23.8}
\newcommand{\GainC}{3.20}
\newcommand{\DutyC}{71}
\newcommand{\RateC}{5554}
\newcommand{\WaitC}{1.1}
\newcommand{\MedC}{0.5}
\newcommand{\PninetyC}{8}
\newcommand{\MedSmallC}{0.05}
\newcommand{\ThrD}{18.3}
\newcommand{\GainD}{2.46}
\newcommand{\DutyD}{59}
\newcommand{\RateD}{2992}
\newcommand{\WaitD}{3.2}
\newcommand{\MedD}{1.2}
\newcommand{\PninetyD}{15}
\newcommand{\MedSmallD}{0.20}
\newcommand{\DoverC}{77}
\newcommand{\MinGainD}{2.34}
\newcommand{\MaxGainD}{2.62}
\newcommand{\NDays}{24}
\newcommand{\CbOne}{1.00}
\newcommand{\CbFour}{1.76}
\newcommand{\CbSixteen}{2.47}
\newcommand{\CbSixtyFour}{2.84}
\newcommand{\CbSixtyFourGenie}{2.91}
\newcommand{\CbSixtyFourFive}{2.75}

\newcommand{\HwFour}{1.66}
\newcommand{\HwEight}{2.16}
\newcommand{\HwTwelve}{2.37}
\newcommand{\HwSixteen}{2.47}
\newcommand{\HwTwenty}{2.49}
\newcommand{\HwOneBit}{2.18}
\newcommand{\HwThreeBit}{2.52}
\newcommand{\HwCont}{2.54}
\newcommand{\HwLossZero}{2.60}
\newcommand{\HwLossThree}{2.21}
\newcommand{\HwEightSide}{27}
\newcommand{\BestHFive}{1}
\newcommand{\BestHSixteen}{6}
\newcommand{\DistFive}{1.99}
\newcommand{\DistThousand}{2.47}
\newcommand{\DistThirteenH}{3.29}
\newcommand{\DistSixteenH}{4.05}
\newcommand{\SensMin}{2.28}
\newcommand{\SensMax}{4.30}

\newcommand{\SensFixed}{2.11}

\newcommand{\JamAMinusTen}{0.02}
\newcommand{\JamDMinusTen}{0.39}
\newcommand{\JamNMinusTen}{1.02}
\newcommand{\JamAZero}{0.00}
\newcommand{\JamDZero}{0.13}
\newcommand{\JamNZero}{0.43}
\newcommand{\JamATen}{0.00}
\newcommand{\JamDTen}{0.03}
\newcommand{\JamNTen}{0.11}

\newcommand{\JamNullDepth}{12}

\title{VHF Reconfigurable Intelligent Surfaces for Meteor Burst Communication}

\author{Rajiv Thummala \footnote{PhD Student, Sibley School of Mechanical and Aerospace Engineering, 124 Hoy Rd, AIAA Member.},
Luke Flores \footnote{Tech Policy Institute Junior Fellow, Cornell Brooks School of Public Policy; Research Assistant, Sibley School of Mechanical and Aerospace Engineering.},
and Gregory Falco \footnote{Assistant Professor, Sibley School of Mechanical and Aerospace Engineering, 124 Hoy Rd, AIAA Member.}}

\affil{Cornell University, Ithaca, NY, 14850}

\begin{document}

\maketitle

\begin{abstract}
Meteor Burst Communication (MBC) utilizes the transient ionized trails left by meteors to reflect Very High Frequency (VHF) signals, enabling long-range, beyond-line-of-sight communication without reliance on terrestrial or satellite infrastructure. Despite its resilience in austere and contested environments, MBC is limited by brief communication windows, low signal-to-noise ratio (SNR), and inconsistent channel availability. Recent advancements in Reconfigurable Intelligent Surfaces (RIS) offer a transformative solution by enabling dynamic control over signal propagation through programmable reflective elements. This paper proposes a novel system architecture that integrates RIS into MBC networks, incorporating real-time adaptive control, optimized deployment strategies, and Monte Carlo simulations to refine RIS placement and operational parameters. 
\end{abstract}

\section*{Nomenclature}

{\renewcommand\arraystretch{1.0}
\noindent\begin{tabular}{@{}l @{\quad=\quad} l@{}}
$C/N_0$ & carrier-to-noise density ratio \\
$D$ & ambipolar diffusion coefficient \\
$E_b/N_0$ & energy-per-bit to noise-density ratio \\
$G_T, G_R$ & master and remote antenna gains \\
$K$ & number of beams in the RIS codebook \\
$N$ & number of RIS elements \\
$P_T, P_R$ & transmitted and received power \\
$q$ & meteor-trail electron line density \\
$q_{\max}$ & peak electron line density \\
$R_T, R_R$ & distances from specular point to master and remote \\
$r_0$ & initial meteor-trail radius \\
$r_e$ & classical electron radius \\
$\mathcal{S}$ & RIS beam codebook \\
$s$ & meteor mass index \\
$t_{\mathrm{acq}}$ & beam-acquisition delay \\
$t_d$ & probe dwell time per beam \\
$t_{\mathrm{hs}}$ & communication handshake time \\
$\beta$ & trail angle relative to the propagation plane \\
$\Gamma$ & combined polarization factor \\
$\eta$ & aperture efficiency \\
$\theta_n$ & reflection phase of RIS element $n$ \\
$\lambda$ & wavelength \\
$\sigma$ & RIS bistatic cross-section \\
$\tau$ & underdense-trail decay time constant \\
$\phi$ & half the forward-scatter angle \\
$\chi$ & meteor-radiant zenith angle \\
\end{tabular}}

\section{Introduction}

Meteor burst communication (MBC) exploits transient ionized trails produced by meteors at altitudes of approximately 80--120~km to support beyond-line-of-sight very high frequency (VHF) communication over paths approaching 2000~km \cite{yavuz}. Unlike satellite or terrestrial relay networks, MBC requires no persistent intermediate infrastructure, making it attractive for remote, degraded, or contested environments. Its principal limitation is the propagation channel itself. Suitable meteor trails occur intermittently, typically remain usable for only fractions of a second to several seconds, and often produce weak received signals \cite{cohen1989meteor}. Consequently, MBC performance is governed by both the frequency with which usable trails occur and the amount of data that can be transferred during each short-lived propagation opportunity.

One approach to improving this performance is to increase antenna gain while retaining sufficient angular coverage to exploit meteor trails distributed across the sky. Reconfigurable intelligent surfaces (RIS), also commonly referred to as intelligent reflecting surfaces (IRS), provide a potential means of achieving this trade-off. These surfaces consist of many passive or nearly passive programmable elements whose reflection phase, and in some implementations amplitude, can be electronically controlled to reshape the reflected field \cite{wu2021intelligent, huang2019reconfigurable}. This enables a large aperture to form and steer high-gain beams without requiring a dedicated radio-frequency chain for each element. In an MBC link, such a surface could therefore steer toward the specular point associated with each usable meteor trail while retaining coverage over the broader region in which those trails occur.

The appropriate placement of an RIS in an MBC link, however, is not obvious. Conventional RIS architectures typically place the surface between a transmitter and receiver to create or strengthen an indirect propagation path. Applying that architecture directly to MBC is problematic because the surface introduces an additional bistatic propagation segment. For a passive RIS located far from both terminals, the resulting path loss scales with the product of the transmitter-to-RIS and RIS-to-receiver distances \cite{tang2021wireless, ozdogan2020intelligent}. At MBC path lengths, this additional loss can overwhelm the aperture gain provided by the surface. A complete link-budget analysis therefore motivates reconsidering the RIS as part of the terminal aperture rather than as a remotely located propagation aid.

This observation motivates a different architecture. Rather than treating the RIS as a remote propagation aid, this work integrates it with the master terminal as a large, electronically steerable reflectarray. In this configuration, the RIS becomes part of the terminal antenna aperture and can direct high gain toward the portions of the sky where usable specular reflections occur. Because those reflection points vary between meteor events, the problem is no longer simply one of maximizing aperture gain; the terminal must determine which beam directions are most useful, acquire a newly formed trail quickly enough to exploit its limited lifetime, and adapt the communication rate as the trail evolves.

This work expands on our preliminary study of RIS-assisted meteor burst communication \cite{thummala2025vhf} by developing the concept in greater detail through a more complete link-budget analysis, a terminal-side RIS architecture, and a broader physics-based simulation and evaluation.

The present study first determines where a passive RIS can provide useful gain in an MBC link and shows that remote placements are severely limited by bistatic path loss. It then develops a terminal-side architecture in which a feed-illuminated reflectarray at the master station provides electronically steerable high gain using a single radio-frequency chain. The control framework combines offline beam-codebook design, millisecond-scale beam-sweep acquisition, and burst-level rate adaptation.

The architecture is evaluated using a GPU-accelerated, physics-based Monte Carlo model of the MBC channel that incorporates sporadic meteor populations, forward-scatter geometry, underdense and overdense trail models, antenna patterns, propagation effects, and receiver noise. The model is calibrated using a single meteor-flux parameter and evaluated against established characteristics of MBC links. Performance is then compared with a conventional Yagi baseline, the same RIS aperture operated with a fixed beam, and an ideal-steering upper bound. The evaluation considers throughput, message-delivery latency, codebook size, acquisition time, panel size, phase resolution, link distance, meteor-model uncertainty, and operation in the presence of airborne jamming.

\section{Motivation}
Meteor burst communication is primarily valuable as a resilient, infrastructure-independent communication method for environments in which conventional terrestrial or satellite links are unavailable, degraded, or undesirable. Its practical utility, however, is constrained by intermittent channel availability and limited burst capacity, making throughput and message-delivery latency central performance considerations.

These limitations affect several established and proposed MBC applications. In defense communications, MBC has been valued for its low probability of intercept and resistance to jamming \cite{jernovics1990meteor}. As both properties depend in part on antenna directivity, an electronically steerable, high-gain aperture could improve link performance while also providing additional spatial selectivity. In environmental monitoring, networks such as SNOTEL have used MBC to collect data from geographically distributed remote stations \cite{cumberland2004understanding}. For such networks, improving the master station is particularly attractive because a single infrastructure upgrade can potentially improve communication with multiple remote terminals without requiring modifications at each site. MBC-like communication architectures have also been proposed for planetary exploration, where communication infrastructure is inherently limited \cite{charania2002networks}.

Across these applications, improving MBC performance requires more than increasing peak received power. Because communication opportunities are transient and irregular, system performance depends on how frequently usable trails can be acquired, how much information can be transmitted during each burst, and how long a message must wait before sufficient channel capacity becomes available. The evaluation in this work therefore emphasizes mean throughput and message-delivery latency, together with the underlying trail availability and channel-usage characteristics that determine them.

\section{Literature Review}
\subsection{Meteor Burst Communication}

Meteor Burst Communication (MBC) exploits the transient ionized trails left by meteors entering Earth's atmosphere to reflect radio signals, enabling long-distance communication. Operating in the Very High Frequency (VHF) spectrum, typically between 30--100 MHz, MBC transmits data in short bursts during the brief periods when suitable meteor trails are available, occurring approximately every 4 to 20 seconds \cite{cohen1989meteor}. The technique offers a low probability of interception, resistance to jamming, and natural time sharing of the channel among stations \cite{cumberland2004understanding}.

MBC supports beyond-line-of-sight (BLOS) communication over distances of up to roughly 2,000 kilometers, making it a valuable tool in environments lacking traditional satellite or terrestrial infrastructure \cite{cannon1987evolution}. Its robustness in austere or contested environments underscores its utility for military long-haul communications \cite{jernovics1990meteor}, remote monitoring systems \cite{cumberland2004understanding}, and even exploratory extraterrestrial applications \cite{charania2002networks}. However, limitations such as low data rates and sporadic meteor trail availability constrain its broader adoption.

Research has sought to address these limitations through adaptive techniques such as variable-rate coding, which significantly enhances throughput \cite{pursley1989variable}, spread-spectrum multiple access to increase the data delivered per trail \cite{aly2001meteor}, and link protocols tailored to rural data-collection networks \cite{whitby1992protocol}. Signal degradation caused by polarization rotation, particularly in linearly polarized systems operating near 40 MHz, is a further known source of loss \cite{cannon1986polarization}. Despite these hurdles, MBC continues to offer a cost-effective and infrastructure-independent communication solution for specific use cases.

The development of MBC dates back to the mid-20th century. The first operational system, JANET, was introduced in Canada in the 1950s, and NATO's COMET (COmmunications by MEteor Trails) system followed in the 1960s, incorporating Automatic Repeat reQuest (ARQ) for error correction \cite{cannon1987evolution}. Although interest in MBC waned with the rise of satellite communications, it experienced renewed attention for high-latitude and high-security applications where satellite solutions proved inadequate \cite{cannon1987evolution}. MBC's inherent security features and minimal infrastructure requirements continue to make it relevant for secure and resilient communication in remote or high-risk environments.

\subsubsection{Applications of Meteor Burst Communication}

Meteor Burst Communication (MBC) demonstrates versatility across multiple sectors, leveraging its unique attributes of long-range transmission, minimal infrastructure dependence, and resilience to environmental conditions.

The military has been a primary beneficiary of MBC technology due to its inherent security features and operational reliability in austere environments. MBC's resistance to jamming and low probability of intercept make it valuable as an additional means of long-haul communication in contested environments \cite{jernovics1990meteor}.

In the civilian domain, MBC is widely used for environmental monitoring. The SNOTEL (SNOwpack TELemetry) system, managed by the U.S. Department of Agriculture, employs MBC to collect hydrological and meteorological data from remote sites across the western United States, supporting water resource management and flood forecasting \cite{cumberland2004understanding}.

MBC's independence from terrestrial infrastructure also makes it attractive for data collection in rural regions without conventional communications coverage \cite{whitby1992protocol}.

With the advent of space exploration, researchers are exploring MBC's potential for interplanetary communication. The idea of using meteor trails in extraterrestrial atmospheres presents a novel approach for maintaining communication with planetary missions, for example on Mars \cite{charania2002networks}.

In summary, MBC's various applications underscore its enduring relevance across multiple domains. Ongoing advancements in data rates and system reliability are poised to expand its utility, particularly in scenarios where traditional communication methods face inherent limitations.

\subsection{Reconfigurable Intelligent Surfaces}

Reconfigurable Intelligent Surfaces (RIS) have emerged as a transformative technology in wireless communication, particularly for advancing next-generation networks. RIS, also referred to as Intelligent Reflecting Surfaces (IRS), are programmable structures consisting of numerous low-cost, tunable reflecting elements that impose controllable phase shifts on incident signals, thereby enhancing propagation between transmitters and receivers \cite{wu2021intelligent}. An RIS operates passively: it neither amplifies nor demodulates the signal, but reflects the incident wave with a phase profile that is set electronically, typically by switching PIN or varactor diodes in each element \cite{huang2019reconfigurable}. This enables RIS to reconfigure the wireless environment, creating what is known as a smart radio environment, which improves signal strength, coverage, and reliability \cite{wu2021intelligent}.

\subsubsection{Technological Foundations and Key Developments}

The concept of RIS emerged as a solution to overcome the inherent limitations of traditional wireless communication systems, where the propagation medium between transmitters and receivers was historically considered a random entity that degraded signal quality. RIS aims to change this paradigm by granting network operators control over the reflection characteristics of radio waves, enabling propagation channels tailored to specific communication needs \cite{wu2021intelligent}.

Studies have highlighted the broad potential of RIS. Passive RIS beamforming has been shown to improve the energy efficiency of wireless links substantially compared with conventional relaying \cite{huang2019reconfigurable}. RIS have been shown to raise the achievable secrecy rate of multiple-antenna links, strengthening physical-layer security \cite{chu2021secrecy}. Compact user-specific RIS have been proposed to provide the benefits of large antenna arrays on the user side for uplink transmission \cite{liu2022compact}, and RIS carried by aerial platforms have been proposed to extend coverage in 5G and beyond \cite{alfattani2021aerial}. Measurements and physical models further show that an RIS far from both terminals behaves as a bistatic scatterer whose path loss grows with the product of its distances to the transmitter and receiver \cite{tang2021wireless, ozdogan2020intelligent}.

\subsubsection{Applications and Challenges}

The applications of RIS span coverage extension around blockages, capacity enhancement, physical-layer security, and support for unmanned aerial vehicle and Internet-of-Things links, with most work targeting centimeter- and millimeter-wave bands \cite{wu2021intelligent}. Despite its potential, RIS technology faces several challenges that must be addressed to achieve widespread adoption. Accurate channel estimation remains a significant hurdle, particularly in dynamic wireless environments where conditions change rapidly, and real-time optimization of RIS configurations requires efficient algorithms \cite{wu2021intelligent}. Continued research addressing channel estimation, hardware efficiency, and integration will be pivotal in unlocking the full potential of RIS.

\section{Where Can an RIS Help an MBC Link?}\label{sec:placement}

\begin{figure}[t]
\centering
\includegraphics[width=0.6\linewidth]{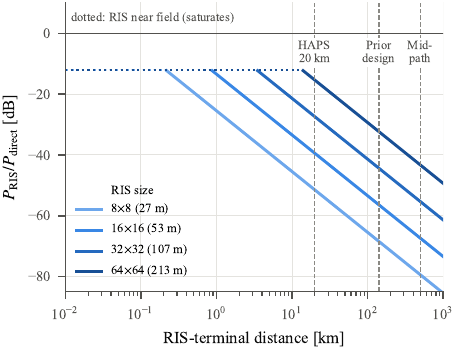}
\caption{Power returned by a passive RIS relative to the direct meteor path, versus the distance between the RIS and the terminal, Eq.~(\ref{eq:placement}). Dotted segments mark the RIS near field, where the ratio saturates.}
\label{fig:placement}
\end{figure}

Consider a passive RIS of area $A$ and aperture efficiency $\eta$ at distance $d$ from a terminal, illuminated by the wave scattered from a meteor trail, and configured to redirect that wave to the terminal. The meteor lies hundreds of kilometers away, so the incident flux density at the RIS and at the terminal is essentially the same. In the RIS far field its peak bistatic cross-section is $\sigma = 4\pi(\eta A)^2/\lambda^2$, so the power it adds relative to the power the terminal already receives directly from the trail is
\begin{equation}
\frac{P_{\mathrm{RIS}}}{P_{\mathrm{direct}}} = \frac{\sigma}{4\pi d^2} = \left(\frac{\eta A}{\lambda d}\right)^2 .
\label{eq:placement}
\end{equation}
Within the RIS near field, $d < 2D^2/\lambda$ for a panel of side $D$, the ratio saturates below 0~dB, the limit of a plane mirror. Figure~\ref{fig:placement} evaluates Eq.~(\ref{eq:placement}) at 45~MHz with $\eta=0.5$. A $16\times16$ surface (53~m square) is \PlaceHAPS~dB below the direct path on a HAPS at 20~km, \PlaceOld~dB below at the position used in the earlier version of this study, and \PlaceMid~dB below at mid-path. Breaking even on a HAPS would require a surface about \HAPSSide~m on a side. A separate passive RIS near the terminal adds at most a path comparable to the direct one, less than 3~dB.

The same surface is useful only when it becomes the terminal's own aperture, a reflectarray antenna fed by an antenna a few tens of meters away, it concentrates the energy that the feed alone would spread over a wide beam. The remainder of the paper therefore studies a terminal-side RIS.

\section{Terminal-Side RIS Architecture}\label{sec:arch}

\subsection{Hardware}

\begin{figure}
    \centering
    \includegraphics[width=0.75\linewidth]{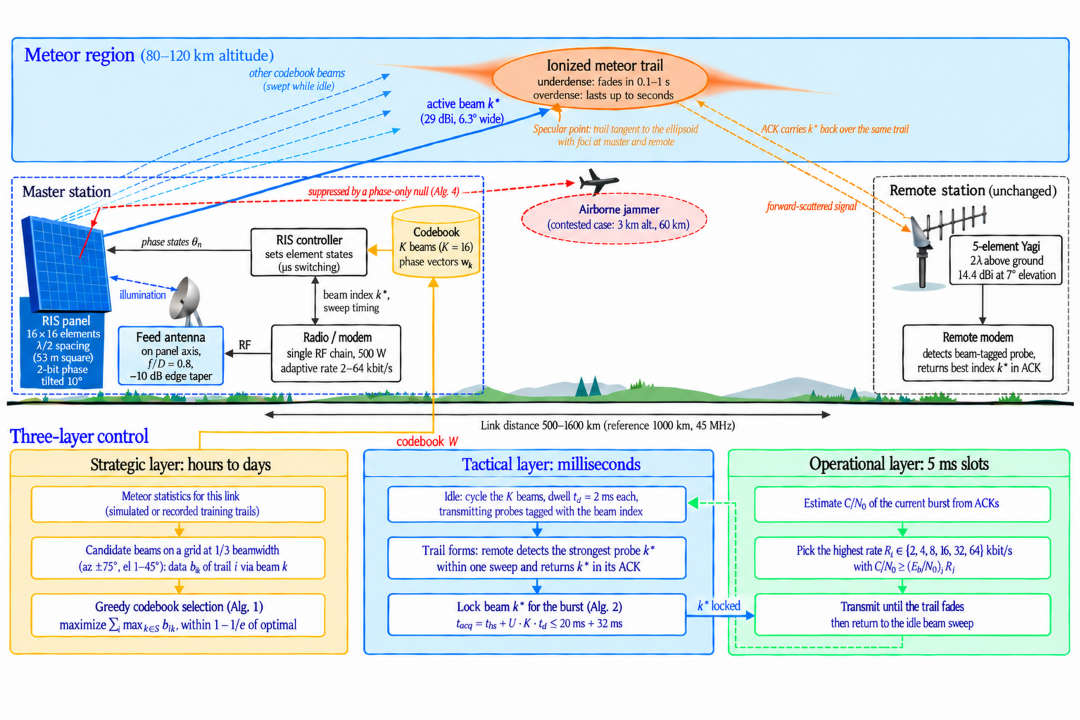}
    \caption{Terminal-side RIS architecture and control framework for meteor
burst communication. The master station uses a single-feed RIS to steer
among a predefined codebook of beams toward candidate meteor-scatter
regions. The strategic layer selects the beam codebook from link-specific
meteor statistics, the tactical layer performs millisecond-scale beam
sweeping and trail acquisition, and the operational layer adapts the
transmission rate during each usable burst. The remote station retains a
conventional Yagi antenna. The contested-operation case additionally
applies phase-only null steering toward an airborne jammer.}
    \label{fig:architecture}
\end{figure}

The proposed terminal-side architecture is shown in
Fig.~\ref{fig:architecture}. The master station of an MBC network uses
the RIS as a reconfigurable reflectarray. A vertical panel of
$N_x\times N_z$ elements at half-wavelength spacing (3.33~m at
45~MHz) stands on the ground, tilted back by $10^\circ$ and facing
along the great-circle path. A small feed antenna on the panel axis,
at a focal distance of 0.8 times the panel side, illuminates it with
a $-10$~dB edge taper. Each element has a 2-bit phase state. The
single feed connects to the station's existing radio, so the RIS
requires only one radio-frequency chain. The remote station retains
a conventional Yagi antenna.

\begin{figure}[t]
\centering
\begin{tikzpicture}[scale=0.9, every node/.style={font=\footnotesize}]
  \draw[thick] (-0.3,0) -- (9.3,0);
  \fill[pattern=north east lines] (-0.3,0) rectangle (9.3,-0.12);
  \draw[line width=2.2pt, blue!60!black] (0.2,0.15) -- (0.55,2.2);
  \node[align=center, anchor=north] at (0.4,-0.2) {RIS panel\\($N_x\times N_z$, 2-bit)};
  \draw[thick] (1.9,1.05) -- (1.9,0);
  \fill (1.9,1.2) circle (0.07);
  \node[anchor=west] at (2.0,1.25) {feed};
  \draw[-{Stealth}, gray] (1.85,1.2) -- (0.55,1.5);
  \draw[-{Stealth}, gray] (1.85,1.2) -- (0.45,0.8);
  \draw[-{Stealth}, blue!60!black, thick] (0.6,1.6) -- (4.0,3.6);
  \draw[-{Stealth}, blue!30, thick, dashed] (0.6,1.6) -- (3.6,2.7);
  \draw[-{Stealth}, blue!30, thick, dashed] (0.6,1.6) -- (3.2,4.1);
  \node[anchor=south east, align=right] at (2.75,3.1) {codebook\\beams};
  \draw[decorate, decoration={snake, amplitude=0.6pt, segment length=4pt}, orange!80!black, line width=1.2pt] (3.8,4.5) -- (5.6,3.4);
  \node[anchor=south west, align=left] at (5.0,3.9) {ionized trail\\(80--120 km)};
  \draw[dashed, gray] (0.6,1.6) .. controls (4.6,5.6) .. (8.6,0.9);
  \node[anchor=south, gray] at (7.8,2.2) {specular ellipsoid};
  \draw[thick] (8.6,0) -- (8.6,0.9);
  \draw[thick] (8.3,0.9) -- (8.9,0.9);
  \draw[thick] (8.4,0.8) -- (8.4,1.0); \draw[thick] (8.6,0.8) -- (8.6,1.0); \draw[thick] (8.8,0.8) -- (8.8,1.0);
  \node[anchor=north] at (8.6,-0.2) {remote (Yagi)};
  \draw[-{Stealth}, orange!80!black] (4.7,3.95) -- (8.5,1.0);
  \node[anchor=north] at (4.7,-0.2) {1000 km};
  \draw[{Stealth}-{Stealth}, gray] (1.0,-0.95) -- (8.3,-0.95);
\end{tikzpicture}
\caption{Terminal-side RIS: the master's radio illuminates a large ground-mounted RIS through a single feed. The RIS points one of $K$ codebook beams at the specular point of the current trail; the remote uses a conventional Yagi.}
\label{fig:concept}
\end{figure}
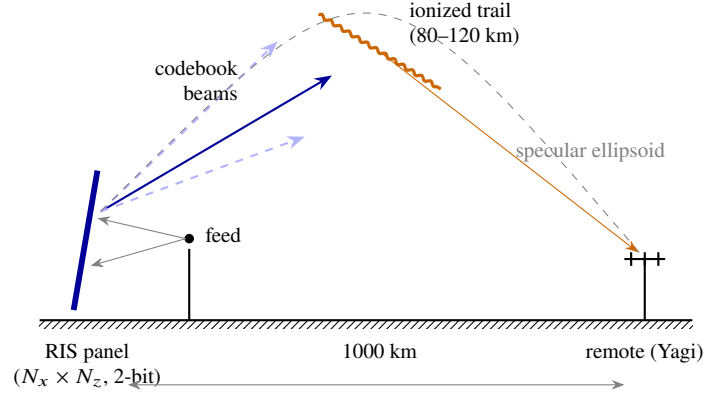

A $16\times16$ panel is 53~m square, comparable to the curtain arrays of high-frequency broadcast stations and to the rhombic antennas historically used at MBC masters. Section~\ref{sec:hw} shows how the benefit scales down to smaller panels. Figure~\ref{fig:concept} illustrates the resulting terminal-side configuration, in which the master RIS directs a selected codebook beam toward the meteor-trail specular point while the remote retains a conventional Yagi antenna.

\subsection{Three-Layer Control}\label{sec:control}

The control problem separates cleanly by time scale because an RIS element switches in microseconds while a trail lasts hundreds of milliseconds.

Strategic layer (hours to days) covers codebook design. Where usable trails can occur for a given link is fixed by the link geometry and the meteor radiant distribution. Offline, the master evaluates a dense grid of candidate beams (spaced at one third of the half-power beamwidth over azimuths within $\pm75^\circ$ of the path and elevations of $1^\circ$--$45^\circ$) on simulated or recorded trails. Let $b_{ik}$ be the data trail $i$ would deliver through candidate beam $k$. The codebook $\mathcal{S}$ of $K$ beams maximizes
\begin{equation}
F(\mathcal{S}) = \sum_i \max_{k\in\mathcal{S}} b_{ik},
\label{eq:greedy}
\end{equation}
which is monotone and submodular, so adding beams greedily gives nested codebooks that are within $1-1/e$ of optimal for every $K$ \cite{nemhauser1978analysis}. The first greedy beam is the best single fixed beam and defines configuration B below. Training uses days and random seeds independent of those used for evaluation.
 Algorithm~\ref{alg:codebook} summarizes the procedure.

\begin{algorithm}[t]
\caption{Strategic layer: greedy codebook design}
\label{alg:codebook}
\begin{algorithmic}[1]
\Require candidate beam directions $\{\hat{\mathbf{u}}_k\}_{k=1}^{M}$ on a grid at one third of the half-power beamwidth; training trails $i=1,\dots,I$ from independent days; phase resolution $b$ bits; codebook size $K$
\Ensure ordered codebook $\mathcal{S}=(k_1,\dots,k_K)$ and element phases $\theta_{n,k}$
\For{$k = 1,\dots,M$}
  \State $\theta_{n,k} \gets \mathrm{Q}_b\!\left[-\arg s_n(\hat{\mathbf{u}}_k)\right]$ for all elements $n$ \Comment{conjugate match, quantize}
\EndFor
\For{each trail $i$ and candidate $k$}
  \State $G_{ik} \gets G_{\mathrm{RIS},k}(\hat{\mathbf{u}}_{T,i})\,G_R(\hat{\mathbf{u}}_{R,i})\,\Gamma_i$
  \State $b_{ik} \gets$ data delivered by trail $i$ through beam $k$ with rate adaptation after $t_{\mathrm{hs}}$
\EndFor
\State discard trails with $\max_k b_{ik} = 0$; \quad $c_i \gets 0$ for all $i$; \quad $\mathcal{S} \gets ()$
\For{$m = 1,\dots,K$}
  \State $k_m \gets \arg\max_{k\notin\mathcal{S}} \sum_i \max(c_i, b_{ik})$ \Comment{largest marginal gain of Eq.~(\ref{eq:greedy})}
  \State append $k_m$ to $\mathcal{S}$; \quad $c_i \gets \max(c_i, b_{ik_m})$ for all $i$
\EndFor
\State \Return $\mathcal{S}$ and $\{\theta_{n,k}\}_{k\in\mathcal{S}}$ \Comment{every prefix is a smaller codebook; $k_1$ is configuration B}
\end{algorithmic}
\end{algorithm}

Tactical layer (milliseconds) covers acquisition. While idle, the master transmits probes continuously and cycles the RIS through its $K$ beams with a dwell $t_d$ per beam; each probe carries its beam index, as in the beam-sweep procedures of millimeter-wave cellular systems \cite{giordani2019tutorial}. When a trail forms, the remote detects the probe of the best beam within one sweep and returns that index in its acknowledgement. The acquisition delay is therefore
\begin{equation}
t_{\mathrm{acq}} = t_{\mathrm{hs}} + U\,K\,t_d, \qquad U\sim\mathcal{U}(0,1),
\label{eq:acq}
\end{equation}
where $t_{\mathrm{hs}}=20$~ms is the handshake already required by conventional MBC. With $K=16$ and $t_d=2$~ms, a 32-bit probe at 16~kbit/s, the sweep adds at most 32~ms.

Operational layer (per slot) covers rate adaptation. During a burst, the RIS holds the selected beam (the specular point of a trail does not move appreciably during its life) and the link selects each 5~ms slot's rate from 2, 4, 8, 16, 32, or 64~kbit/s according to the received $C/N_0$ (required $E_b/N_0$ of 6, 6, 6, 6, 7, and 11~dB).

The tactical and operational layers together form the master's online loop, Algorithm~\ref{alg:online}. In the simulation, the sweep is represented by Eq.~(\ref{eq:acq}) with the remote identifying the best beam.

\begin{algorithm}[t]
\caption{Tactical and operational layers: online operation of the master}
\label{alg:online}
\begin{algorithmic}[1]
\Require codebook $\mathcal{S}=(k_1,\dots,k_K)$; probe dwell $t_d$; rates $R_1<\dots<R_J$ with required $(E_b/N_0)_j$
\Loop
  \State $m \gets 1$ \Comment{idle: beam sweep}
  \Repeat
    \State load phases $\theta_{n,k_m}$ into the RIS \Comment{element switching in microseconds}
    \State transmit a probe tagged with index $m$ for $t_d$
    \State $m \gets (m \bmod K) + 1$
  \Until{an acknowledgement carrying index $m^*$ is received}
  \State load phases $\theta_{n,k_{m^*}}$ and complete the handshake ($t_{\mathrm{hs}}$)
  \Repeat \Comment{burst: one iteration per 5~ms slot}
    \State estimate $C/N_0$ from the remote's acknowledgements
    \State $R \gets \max\{R_j : (C/N_0)_{\mathrm{dBHz}} \ge (E_b/N_0)_{j,\mathrm{dB}} + 10\log_{10}R_j\}$
    \State transmit the next packet at rate $R$
  \Until{$(C/N_0)_{\mathrm{dBHz}} < (E_b/N_0)_{1,\mathrm{dB}} + 10\log_{10}R_1$} \Comment{trail has faded}
\EndLoop
\end{algorithmic}
\end{algorithm}

\section{Simulation Method}\label{sec:sim}

\subsection{Meteor Population}

Based on the sporadic-source distributions reported in
\cite{jones1993sporadic,campbellbrown2008high}, the simulator represents
the sporadic meteor population using five sources fixed in sun-centered
ecliptic coordinates: the helion and antihelion sources
($\lambda-\lambda_\odot = 340^\circ$ and $200^\circ$, 30~km/s), the apex
source ($270^\circ$, 55~km/s), and the north and south toroidal sources
($270^\circ$, $\pm58^\circ$ ecliptic latitude, 35~km/s), with weights
0.25, 0.25, 0.30, 0.10, and 0.10 and angular spreads of
$10^\circ$--$15^\circ$. For each meteor, the local solar time and day of
year determine the radiant direction in the horizon frame of the link
midpoint (latitude $45^\circ$, path oriented east--west). The flux through
a horizontal surface is weighted by $\cos\chi$.

The cumulative number of meteors whose peak line density at vertical
incidence exceeds $q$ follows
$N(>q)\propto q^{-(s-1)}$ with $s=2.0$. The model assumes a peak line
density $q_{\max}=q\cos\chi$ and represents the height of maximum
ionization as

\begin{equation}
h_{\max} =
88 + 0.3(v-20)
+ \min[H\ln(1/\cos\chi),\,10]
- 2.5\log_{10}(q_{\max}/10^{13})
+ \epsilon
\quad \text{km},
\end{equation}

where $v$ is the meteor speed in km/s, $H=6$~km is the atmospheric scale
height, and $\epsilon\sim\mathcal{N}(0,3^2)$~km accounts for variability in
the ionization height. Along the trail, the line density is represented by

\[
q(x)=\tfrac{9}{4}q_{\max}x\left(1-\sqrt{x}/3\right)^2,
\]

where $x$ is the ratio of atmospheric pressure to that at $h_{\max}$,
evaluated from ablation onset ($x=0.02$) to burnout ($x=9$).

\subsection{Forward-Scatter Geometry and Scattering}

For each meteor with trail axis $\hat{\mathbf{d}}$, the specular point $\mathbf{X}$ is where $R_T+R_R$ is stationary along the trail,
\begin{equation}
\hat{\mathbf{d}}\cdot\left(\hat{\mathbf{u}}_T+\hat{\mathbf{u}}_R\right)=0,
\end{equation}
with $\hat{\mathbf{u}}_{T,R}$ the unit vectors from the terminals to $\mathbf{X}$. It is found by Newton iteration on a spherical Earth. A trail is retained if the specular point lies within its ionized column and at least $1^\circ$ above both horizons. For underdense trails the received power is 
\begin{equation}
P_R(t) = \frac{P_T G_T G_R \lambda^3 q^2 r_e^2 \, \Gamma}{16\pi^2 R_T R_R (R_T+R_R)(1-\cos^2\beta\sin^2\phi)}\,
\exp\!\left(-\frac{8\pi^2 r_0^2}{\lambda^2\sec^2\phi}\right) \exp\!\left(-\frac{t}{\tau}\right),
\qquad \tau = \frac{\lambda^2\sec^2\phi}{32\pi^2 D},
\end{equation}
where $\Gamma$ combines the polarization factor $\sin^2\gamma$ with the receive polarization mismatch of the horizontally polarized antennas. For overdense trails \cite{hines1957forward}
\begin{equation}
P_R(t) = \frac{P_T G_T G_R \lambda^2 \, \Gamma}{32\pi^2 R_T R_R (R_T+R_R)(1-\cos^2\beta\sin^2\phi)}
\left[\frac{4Dt+r_0^2}{\sec^2\phi}\ln\frac{r_e q\lambda^2\sec^2\phi}{\pi^2(4Dt+r_0^2)}\right]^{1/2},
\end{equation}
until the logarithm vanishes or 15~s elapse, the latter representing destruction by wind shear. The diffusion coefficient and initial radius increase with height as $\log_{10}D = 0.067h-5.6$ and $\log_{10}r_0 = 0.037h-3.55$ ($h$ in km).

\subsection{Antennas, Noise, and Link Layer}

The baseline antenna at both ends is a five-element horizontal Yagi (9~dBi in free space, 15~dB front-to-back ratio) at two wavelengths above perfectly reflecting ground, giving a peak gain of 14.4~dBi at $7.2^\circ$ elevation. The height was chosen to maximize the baseline's throughput on the reference link among mounting heights of 1--4 wavelengths, so that the RIS is compared with a well-sited conventional antenna. The RIS field pattern toward direction $\hat{\mathbf{u}}$ is
\begin{equation}
E(\hat{\mathbf{u}}) = \sum_{n=1}^{N} s_n(\hat{\mathbf{u}})\, e^{j\theta_n}, \qquad
s_n(\hat{\mathbf{u}}) = a_n e^{-jk\ell_n} p(\hat{\mathbf{u}}) \left[g(\hat{\mathbf{u}}) e^{jk\hat{\mathbf{u}}\cdot\mathbf{r}_n} - g'(\hat{\mathbf{u}}) e^{jk\hat{\mathbf{u}}\cdot\mathbf{r}'_n}\right],
\end{equation}
where $a_n$ and $\ell_n$ are the feed illumination and path length to element $n$ at $\mathbf{r}_n$, $\mathbf{r}'_n$ is its ground image, $g$ and $g'$ are cosine element patterns, and $p$ is the dipole polarization factor. A beam toward $\hat{\mathbf{u}}_0$ uses $\theta_n = -\arg s_n(\hat{\mathbf{u}}_0)$ rounded to the nearest 2-bit state. Gain is $G(\hat{\mathbf{u}}) = 4\pi|E|^2/\int_{2\pi}|E|^2\,d\Omega \times \eta_s\eta_L$, with the directivity integral evaluated numerically over the upper hemisphere, a computed spillover efficiency $\eta_s=0.92$, and 1~dB element loss $\eta_L$. The reference $16\times16$ panel reaches 29.2~dBi with a $6.3^\circ$ beamwidth; mutual coupling and feed blockage are neglected. Noise is the ITU-R P.372 median galactic noise, $F_a = 52-23\log_{10}f_{\mathrm{MHz}}$~dB \cite{itur2019p372}, plus a 4~dB receiver noise figure. The master transmits 500~W.

Each simulated day is represented as a timeline of 5~ms slots. Every usable trail becomes available $t_{\mathrm{acq}}$ after it forms and is used at the highest supported rate until it fades; overlapping bursts are combined by taking the best rate in each slot. From the timeline we compute mean throughput, the fraction of time the channel is usable, the wait from a random instant to the next usable slot, and the time to deliver 100~B and 1~kB messages arriving at random instants.

\subsection{GPU Implementation}

The simulator is written in Python with CuPy \cite{okuta2017cupy}. A single fused CUDA kernel draws each candidate meteor from a counter-based random number generator, computes its radiant, ionization, and specular point in single precision, and writes out only the few percent of meteors that produce a specular trail above the line-density floor ($3\times10^{11}$~electrons/m). It processes about $3\times10^9$ candidates per second on an NVIDIA RTX A2000; on identical candidates it agrees with a double-precision reference implementation for all but 8 of 217,433 trails, all within rounding of the line-density floor. RIS gains for all trails and beams are complex matrix products. A simulated day with four antenna configurations takes 4--13~s.
 Algorithm~\ref{alg:sim} summarizes one simulated day.

\begin{algorithm}[t]
\caption{Monte Carlo simulation of one day}
\label{alg:sim}
\begin{algorithmic}[1]
\Require link geometry; day of year; calibrated flux; configurations $\mathcal{C}$
\State $N \sim \mathrm{Poisson}(\text{flux} \times \text{area} \times 24~\mathrm{h})$ candidate meteors
\ForAll{candidates in parallel} \Comment{fused CUDA kernel, one thread per candidate}
  \State draw local time, position, sporadic source, and radiant; accept with probability $\cos\chi$
  \State draw speed $v$, line density $q_{\max}=q\cos\chi$, and height of maximum ionization $h_{\max}$
  \State solve $\hat{\mathbf{d}}\cdot(\hat{\mathbf{u}}_T+\hat{\mathbf{u}}_R)=0$ for the specular point by Newton iteration
  \If{converged, ionized there, above both horizons, and $q \ge q_{\mathrm{floor}}$}
    \State append the trail to the output list (atomic counter)
  \EndIf
\EndFor
\ForAll{configurations $c\in\mathcal{C}$} \Comment{identical trails for every configuration}
  \State best beam and gain per trail: $G = G_T G_R \Gamma$ (RIS beams as complex matrix products)
  \State acquisition delay per trail from Eq.~(\ref{eq:acq}) (sweep) or $t_{\mathrm{hs}}$ (fixed beam)
  \State per 5~ms slot, rate $\gets$ best rate supported by any active burst (GPU scatter-max)
  \State record throughput, usable time, wait for the channel, and message delivery times
\EndFor
\end{algorithmic}
\end{algorithm}

\subsection{Calibration and Validation}\label{sec:validation}

\begin{figure}[t]
\centering
\includegraphics[width=\linewidth]{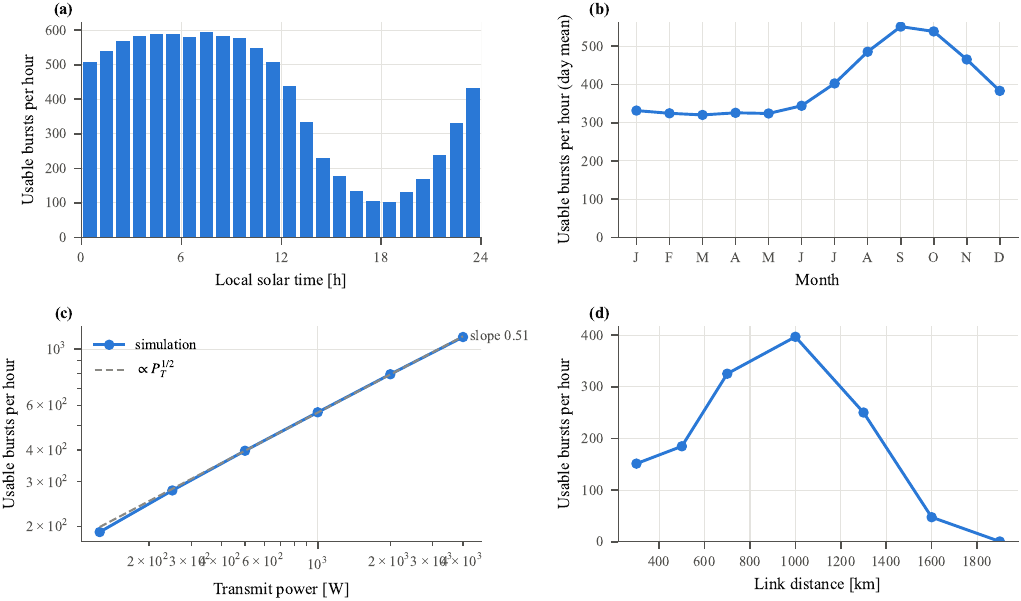}
\caption{Emergent behavior of the calibrated baseline: (a) usable bursts by local time, (b) by month, (c) versus transmitter power, (d) versus link distance.}
\label{fig:validation}
\end{figure}

\begin{figure}[t]
\centering
\includegraphics[width=0.6\linewidth]{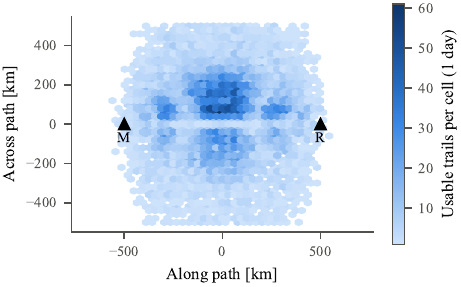}
\caption{Specular points of usable trails for the baseline link over one day. M: master, R: remote.}
\label{fig:hotspots}
\end{figure}

The model has one free constant, the flux of meteors with $q\ge10^{13}$~electrons/m at vertical incidence. It is set so that a legacy link (1000~km, 45~MHz, 500~W, Yagis, fixed 4~kbit/s at $E_b/N_0=8$~dB) sees a usable burst every \TargetInterval~s on the annual average, the middle of the 4--20~s range reported by the NTIA \cite{cohen1989meteor}; the result is \FluxRef~meteors~km$^{-2}$~h$^{-1}$. Nothing else was tuned. Figure~\ref{fig:validation} and Table~\ref{tab:validation} compare emergent behavior with established properties of MBC links. The simulator reproduces the dawn maximum and dusk minimum of meteor rates, the $P_T^{1/2}$ scaling of usable-burst rate with transmitter power, an optimum link distance near 1000~km, and the two off-path hot spots (Fig.~\ref{fig:hotspots}).

Two discrepancies bound the absolute accuracy. The seasonal variation (\SeasonRatio:1) is weaker than the 2:1--4:1 observed, because the source strengths are held constant through the year; all results are therefore annual averages. The legacy link is usable a larger fraction of the time than typically reported, because long-lived overdense trails dominate its on-time. Absolute throughputs should therefore be read as optimistic, and the paper's conclusions rest on ratios between configurations evaluated on identical meteors. Section~\ref{sec:sens} shows that those ratios are robust to the overdense lifetime, mass index, diffusion rate, and rate mode.

\begin{table}[t]
\centering
\caption{Emergent behavior of the calibrated simulator (legacy baseline link)}
\label{tab:validation}
\begin{tabular}{lll}
\toprule
Property & Simulation & Established behavior \\
\midrule
Diurnal variation & max \DiurnalPeak~h, min \DiurnalMin~h, \DiurnalRatio:1 & dawn maximum, several-fold \\
Power scaling of burst rate & $\propto P_T^{\PowerExp}$ & $\propto P_T^{1/2}$ \\
Most effective link distance & \BestDist~km & $\sim$800--1200~km \\
Specular hot spots & two lobes, median offset \HotspotOffset~km & two lobes off the path \cite{eshleman1957directional} \\
Seasonal variation & \SeasonRatio:1 & 2:1--4:1 (not reproduced) \\
\bottomrule
\end{tabular}
\end{table}

\section{Results}\label{sec:results}

\subsection{Configurations}

All configurations share the remote Yagi and differ at the master:
\begin{description}
\item[A] Yagi (baseline).
\item[B] $16\times16$ 2-bit RIS panel with the best single fixed beam, i.e.\ the same aperture used as a conventional high-gain antenna.
\item[C] The same panel with continuous phases, steered instantly to the best of all 1491 candidate beams for each trail (upper bound).
\item[D] The same panel with 2-bit phases, a 16-beam greedy codebook, and beam-sweep acquisition at 2~ms per beam.
\end{description}
Configurations are compared on identical meteors (common random numbers) over \NDays{} simulated days, two realizations of each mid-month day.

\subsection{Why Steering Matters}

\begin{figure*}[t]
\centering
\includegraphics[width=\linewidth]{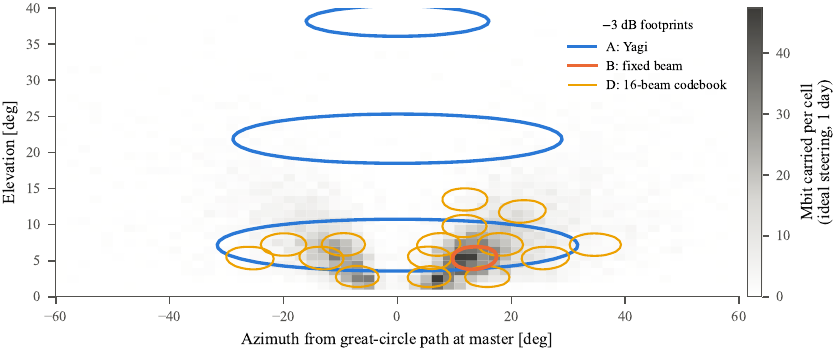}
\caption{Master's view of usable specular points on one day, weighted by the data they carry under ideal steering, with the $-3$~dB footprints of the Yagi, the fixed beam, and the 16 codebook beams.}
\label{fig:skyview}
\end{figure*}

Figure~\ref{fig:skyview} shows, from the master, where the usable specular points lie and how much data they carry. They spread over about $60^\circ$ of azimuth, mostly between $2^\circ$ and $10^\circ$ elevation, in two lobes that correspond to the two hot spots. The Yagi's broad beam covers them at modest gain. A fixed high-gain beam (B) covers only part of one lobe, and which lobe is productive changes with the time of day. The 16-beam codebook (D) tiles the productive region with high-gain beams.

\subsection{Main Comparison}

\begin{figure*}[t]
\centering
\includegraphics[width=\linewidth]{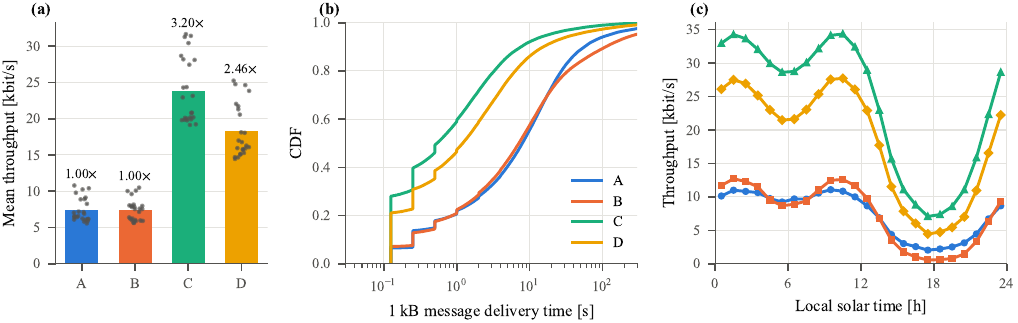}
\caption{Reference link: (a) mean throughput with daily values (dots), (b) distribution of 1~kB message delivery time, (c) throughput by local time.}
\label{fig:main}
\end{figure*}

Table~\ref{tab:main} and Fig.~\ref{fig:main} summarize the reference link. The terminal-side RIS with a practical 16-beam codebook (D) delivers \GainD{} times the baseline throughput and reaches \DoverC\% of the ideal-steering bound (C). The daily gain ranged from \MinGainD{} to \MaxGainD. The improvement comes from using many more trails: D uses \RateD{} trails per hour against \RateA{} for the Yagi, because the extra gain makes weaker, more numerous underdense trails usable. As a result, the channel is idle much less often. The median 1~kB message is delivered in \MedD~s instead of \MedA~s, and the 90th percentile falls from \PninetyA~s to \PninetyD~s. The same aperture with a fixed beam (B) matches the baseline's throughput (factor \GainB) but not its availability: its average wait for the channel is \WaitB~s against \WaitA~s, because it depends on a single hot spot whose activity varies through the day (Fig.~\ref{fig:main}c). The benefit of the RIS comes from steering, not from aperture alone.

\begin{table}[t]
\centering
\caption{Reference link: 1000~km, 45~MHz, 500~W, adaptive rate, annual average}
\label{tab:main}
\begin{tabular}{lcccc}
\toprule
 & A: Yagi & B: fixed beam & C: ideal RIS & D: codebook RIS \\
\midrule
Mean throughput, kbit/s & \ThrA & \ThrB & \ThrC & \ThrD \\
Relative to A & 1.00 & \GainB & \GainC & \GainD \\
Usable trails per hour & \RateA & \RateB & \RateC & \RateD \\
Channel usable, \% of time & \DutyA & \DutyB & \DutyC & \DutyD \\
Mean wait for channel, s & \WaitA & \WaitB & \WaitC & \WaitD \\
100~B delivery, median, s & \MedSmallA & \MedSmallB & \MedSmallC & \MedSmallD \\
1~kB delivery, median, s & \MedA & \MedB & \MedC & \MedD \\
1~kB delivery, 90th percentile, s & \PninetyA & \PninetyB & \PninetyC & \PninetyD \\
\bottomrule
\end{tabular}
\end{table}

\subsection{Codebook Size and Acquisition Time}

\begin{figure}[t]
\centering
\includegraphics[width=0.6\linewidth]{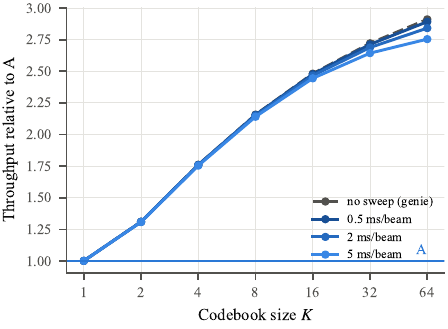}
\caption{Throughput of the codebook RIS relative to the baseline versus codebook size, for several probe dwell times per beam.}
\label{fig:codebook}
\end{figure}

Figure~\ref{fig:codebook} shows the tactical trade-off. With instantaneous (genie) acquisition, throughput rises steadily with $K$ as the codebook covers more of the productive sky with high-gain beams. Beam sweeping costs up to $K t_d$ per trail, which matters most for short underdense bursts, but the cost is small: at 2~ms per beam the gain grows from \CbOne{} at $K=1$ to \CbFour{} at $K=4$, \CbSixteen{} at $K=16$, and \CbSixtyFour{} at $K=64$, against \CbSixtyFourGenie{} with genie acquisition, and even at 5~ms per beam a 64-beam codebook achieves \CbSixtyFourFive. Larger codebooks are therefore worthwhile; the reference design uses $K=16$ as a conservative choice.

\subsection{Panel Size, Phase Resolution, and Element Loss}\label{sec:hw}

\begin{figure}[t]
\centering
\includegraphics[width=\linewidth]{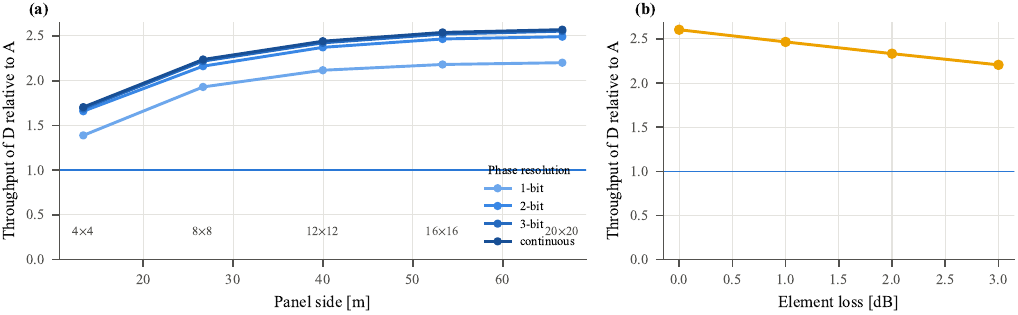}
\caption{Throughput of the codebook RIS relative to the baseline: (a) versus panel size and phase resolution (1~dB element loss), (b) versus element loss for the $16\times16$, 2-bit panel.}
\label{fig:hardware}
\end{figure}

Figure~\ref{fig:hardware} shows how the benefit scales with hardware. Every panel keeps a 16-beam codebook trained for its own beamwidth. A $4\times4$ panel (13~m), whose gain is comparable to the Yagi's, gains a factor of \HwFour{}; an $8\times8$ panel (\HwEightSide~m) \HwEight; $12\times12$ \HwTwelve; $16\times16$ \HwSixteen; and $20\times20$ \HwTwenty. With 16 beams the largest panels begin to leave parts of the sky uncovered, so larger codebooks or faster sweeps become necessary. Phase resolution matters mainly at 1 bit (factor \HwOneBit{} against \HwThreeBit{} at 3 bits and \HwCont{} with continuous phase), consistent with the 3~dB loss and quantization lobes of 1-bit surfaces. Element loss from 0 to 3~dB reduces the gain from \HwLossZero{} to \HwLossThree.

\subsection{Link Distance}

\begin{figure}[t]
\centering
\includegraphics[width=\linewidth]{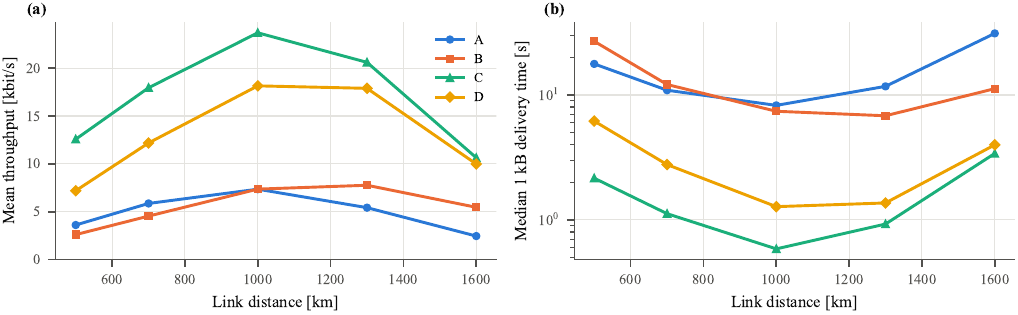}
\caption{(a) Mean throughput and (b) median 1~kB delivery time versus link distance.}
\label{fig:distance}
\end{figure}

Figure~\ref{fig:distance} repeats the comparison from 500 to 1600~km. For each distance, the Yagi height at both ends is first re-optimized over 1--6 wavelengths (from \BestHFive{} wavelength, the lowest tested, at 500~km to \BestHSixteen{} wavelengths at 1600~km), and the codebook is retrained for the link. The relative gain of D over A is \DistFive{} at 500~km, \DistThousand{} at 1000~km, \DistThirteenH{} at 1300~km, and \DistSixteenH{} at 1600~km. On long links the specular points lie within a few degrees of the horizon, where the ground reflection limits a Yagi's gain even at its best height, while the tall RIS aperture still forms a high-gain beam; the relative benefit of the RIS therefore grows with distance.

\subsection{Contested Operation}

\begin{figure}[t]
\centering
\includegraphics[width=0.6\linewidth]{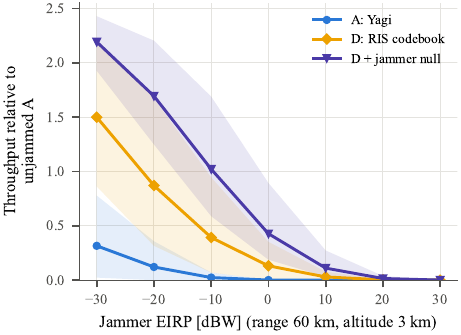}
\caption{Throughput under an airborne noise jammer relative to the unjammed baseline (mean and 10th--90th percentile over jammer azimuths).}
\label{fig:jamming}
\end{figure}

A steerable aperture also changes MBC's behavior under jamming. We place a broadband noise jammer on an aircraft at 3~km altitude, 60~km from the master, at a random azimuth in the forward half-space (eight realizations), and evaluate the master's receiver. Airborne jammers are the relevant threat: a ground jammer beyond the radio horizon is attenuated by terrain, and one within it arrives at grazing incidence, where the ground reflection cancels horizontally polarized reception. Configuration D-null recomputes each codebook beam by phase-only coordinate descent over the 2-bit states to minimize the ratio of its response toward the jammer to that toward its target direction, which requires only the jammer's bearing
 (Algorithm~\ref{alg:null}; four passes are used). To avoid crediting unrealistically deep nulls, every RIS beam, with or without a null, is evaluated with random element errors of $5^\circ$ in phase and 0.5~dB in amplitude. Figure~\ref{fig:jamming} shows throughput relative to the unjammed baseline, averaged over jammer azimuths. At $-10$~dBW jammer EIRP the Yagi delivers \JamAMinusTen{} times its unjammed throughput, the codebook RIS \JamDMinusTen, and the nulling RIS \JamNMinusTen; at 0~dBW the values are \JamAZero, \JamDZero, and \JamNZero; at 10~dBW they are \JamATen, \JamDTen, and \JamNTen. With element errors the median null depth, relative to the strongest codebook beam's response toward the jammer, is \JamNullDepth~dB. As meteor-scattered signals are extremely weak, even a modest airborne jammer overwhelms a conventional terminal. The steerable aperture alone tolerates roughly 10~dB more jammer power than the Yagi for the same throughput, and the null adds roughly another 10~dB.

\begin{algorithm}[t]
\caption{Phase-only null steering for one codebook beam}
\label{alg:null}
\begin{algorithmic}[1]
\Require beam phases $w_n = e^{j\theta_{n,k}}$; target direction $\hat{\mathbf{u}}_k$; jammer bearing $\hat{\mathbf{u}}_J$; phase states $\mathcal{Q}=\{e^{j2\pi q/2^b}\}$; number of passes $P$
\State $a_n \gets s_n(\hat{\mathbf{u}}_J)$, \quad $t_n \gets s_n(\hat{\mathbf{u}}_k)$
\State $A \gets \sum_n a_n w_n$ \Comment{response toward the jammer}
\State $S \gets \sum_n t_n w_n$ \Comment{response toward the target}
\For{$p = 1,\dots,P$}
  \For{each element $n$ in random order}
    \State $w' \gets \arg\min_{x\in\mathcal{Q}} |A + a_n(x-w_n)|^2 \,/\, |S + t_n(x-w_n)|^2$
    \State $A \gets A + a_n(w'-w_n)$; \quad $S \gets S + t_n(w'-w_n)$; \quad $w_n \gets w'$
  \EndFor
\EndFor
\State \Return $\{w_n\}$
\end{algorithmic}
\end{algorithm}

\subsection{Sensitivity to the Meteor Model}\label{sec:sens}

\begin{figure}[t]
\centering
\includegraphics[width=0.6\linewidth]{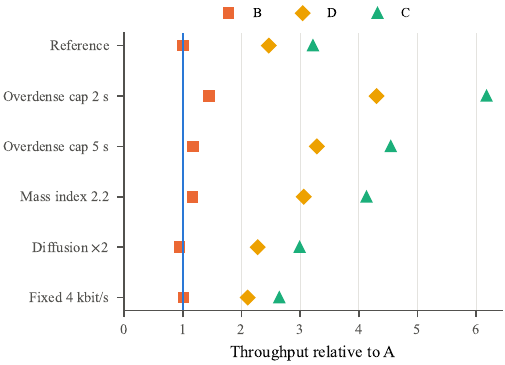}
\caption{Throughput relative to the baseline under alternative meteor models and rate modes; each meteor model is recalibrated to the same legacy burst interval.}
\label{fig:sensitivity}
\end{figure}

The main uncertainty in the channel model is the overdense population, which dominates the baseline's on-time. Figure~\ref{fig:sensitivity} recalibrates the flux for each alternative model so that the legacy link again sees a burst every \TargetInterval~s, then repeats the comparison. Capping overdense lifetimes at 2~s instead of 15~s, raising the mass index to 2.2, or doubling the diffusion coefficient changes the relative throughput of D between \SensMin{} and \SensMax. Models with fewer or shorter overdense trails favor the RIS more, because its advantage comes mostly from the more numerous weak trails. With a fixed 4~kbit/s rate instead of rate adaptation, the gain of D is \SensFixed.

\section{Discussion and Limitations}

The results indicate that the primary benefit of an RIS in meteor burst communication arises when the surface is integrated into the terminal antenna rather than deployed as a separate passive reflector. At the master station, an electronically steered RIS can combine the gain of a large aperture with the angular coverage required to exploit meteor trails distributed across the observable sky. The comparison between the fixed-beam and steerable configurations further shows that aperture size alone is insufficient: the performance improvement depends on the ability to redirect the high-gain beam between successive specular points. The proposed finite codebook and millisecond-scale acquisition procedure recover most of the performance of ideal steering while remaining compatible with the short duration of meteor-scatter events. In contrast, passive RIS placements away from the terminals incur sufficient bistatic path loss that they provide no useful link advantage under the configurations considered here.

The quantitative results should nevertheless be interpreted primarily as comparisons between antenna configurations rather than as predictions of absolute MBC capacity. As discussed in Sec.~\ref{sec:validation}, the simulator underestimates the observed seasonal variation in meteor activity and predicts a larger channel duty cycle than is typically reported. These effects make the absolute throughput estimates optimistic. The relative comparisons are less sensitive to these uncertainties because all configurations are evaluated using identical meteor realizations, and Sec.~\ref{sec:sens} shows that the principal RIS gains persist under changes to the overdense-trail lifetime, meteor mass index, diffusion coefficient, and communication-rate model.

Additional limitations arise from the electromagnetic model of the RIS. The present analysis neglects mutual coupling between elements, feed blockage, structural scattering, and the frequency dependence of the element phase response across the communication bandwidth. The ground is also modeled as flat and perfectly conducting. These assumptions allow the study to isolate the effect of aperture size and electronic steering, but a physical implementation would require full-wave or measured characterization of the reflectarray, its feed, the local ground environment, and associated losses. Similarly, overdense meteor trails are modeled as smooth reflecting structures. Real trails can be distorted by winds and develop fading or fragmentation that is not represented explicitly, although burst-level rate adaptation provides some tolerance to time-varying received power.

The architecture evaluated here also places the RIS only at the master terminal. Adding a steerable aperture at the remote terminal could provide additional link gain, but it would introduce a coupled two-sided acquisition problem and greater terminal complexity. The physical scale of the aperture is another practical constraint. At 45~MHz, the reference $16\times16$ half-wavelength-spaced surface is approximately 53~m on a side. The results in Sec.~\ref{sec:hw}, however, show that smaller $8\times8$ and $12\times12$ surfaces retain a substantial portion of the achievable improvement. A conventional phased array with a comparable aperture could provide similar beamforming capability, but would generally require substantially more active radio-frequency hardware than the single-feed reflectarray architecture considered here.

Future work should therefore focus on experimental validation and reduction of the remaining modeling uncertainty. A first step is a hardware demonstration using a smaller VHF reflectarray to validate the predicted gain, beam-steering behavior, switching time, and achievable null depth. Measurements from an operational MBC link could then be used to characterize the spatial distribution of useful specular points and to train or update the beam codebook from empirical data. A further extension is to evaluate multi-terminal networks in which a single master RIS dynamically serves remote stations located along different paths.

\section{Conclusion}

This work evaluates where and how a reconfigurable intelligent surface can improve meteor burst communication. The analysis shows that a passive RIS placed on a high-altitude platform or elsewhere along the propagation path is ineffective because the additional bistatic propagation loss outweighs the gain provided by the surface. The useful configuration instead integrates the RIS with the master terminal as a feed-illuminated, electronically steerable reflectarray.

In this configuration, electronic steering addresses the fundamental trade-off between antenna gain and angular coverage in MBC. For the 1000~km reference link, a 2-bit $16\times16$ RIS using a 16-beam codebook and 2~ms beam dwell increased mean throughput by a factor of \GainD{} relative to the optimized Yagi baseline and reduced the median delivery time of a 1~kB message from \MedA~s to \MedD~s. The improvement persisted across the meteor-model sensitivity cases considered in this study, while phase-only null steering provided additional resilience in the airborne-jamming scenario. These results indicate that terminal-integrated, electronically steerable apertures provide a practical means of improving the utilization of short-lived meteor-scatter opportunities without modifying the remote terminals.

\bibliography{sampleRight}

\end{document}